\documentclass[12pt]{article}
\usepackage{amsfonts,amssymb,amsmath}
\usepackage[dvips]{epsfig}
\begin{document}
\title{\bf Effect of two-qutrit entanglement on quantum speed limit time of a bipartite V-type open system}
\author{ N. Behzadi $^{a}$
\thanks{E-mail:n.behzadi@tabrizu.ac.ir}  ,
B. Ahansaz $^{b}$,
A. Ektesabi $^{b}$ and
E. Faizi $^{b}$
\\ $^a${\small Research Institute for Fundamental Sciences, University of Tabriz, Iran,}
\\ $^b${\small Physics Department, Azarbaijan shahid madani university, Tabriz, Iran.}} \maketitle

\begin{abstract}
\noindent
In the present paper, quantum speed limit (QSL) time of a bipartite V-type three-level atomic system under the effect of two-qutrit entanglement is investigated. Each party interacts with own independent reservoir. By considering two local unitarily equivalent Wener states and the Horodecki PPT state, as initial states, the QSL time is evaluated for each of them in the respective entangled regions. It is counterintuitively observe that the effect of entanglement on the QSL time driven from each of the initial Werner states are completely different when the degree of non-Markovianity is considerable. In addition, it is interesting that the effect of entanglement of the non-equivalent Horodecki state on the calculated QSL time displays an intermediate behavior relative to the cases obtained for the Werner states.
\\
\\
{\bf PACS Nos:}
\end{abstract}

\section{Introduction}
Quantum speed limit (QSL) is of particular interest and
has attracted much attention in tremendous areas of quantum
physics and quantum information, such as nonequilibrium thermodynamics $\cite{Lutz}$, quantum metrology [2-5],
quantum optimal control [6-11], quantum computation [12-14], and quantum communication $\cite{Lutz,Yung}$.
In these fields, QSL time is a key concept, defined as
the minimum evolution time in which a system evolves from an initial state to a target state.
In the other hand, the QSL time sets a bound on the minimal time a system which needs to evolve
between two distinguishable states, and it can be understood as a generalization of the time-energy uncertainty principle.
For isolated systems, the QSL time under unitary evolution is determined by the
Mandelstam-Tamm bound $\cite{Mandelstam}$ and Margolus-Levitin bound $\cite{Margolus}$.
Because of the inevitable interactions with the environments, quantum systems should be generally regarded
as open systems. Therefore, it is necessary to determine a QSL time for open systems.
Recently, Taddei $et$ $al.$ developed a method to investigate the QSL problem in open systems described by positive non-unitary
maps by using of quantum Fisher information for time estimation $\cite{Taddei}$.
While, for the case of the initial mixed states, a Hermitian operator
is required to minimize the Fisher information in the extended system environment space, which is generally a hard task. Afterwards, Campo $et$ $al.$ exploited the concept of relative purity to derive an
analytical and computable QSL time for open systems undergoing a
completely positive and trace preserving evolution $\cite{Campo}$.
It should be noted that relative entropy can perfectly make a distinction between
an initial pure state and its target state, however it may fail to distinguish an initial mixed state to its target state.
Also, a tight bound on the minimal evolution time of an arbitrarily driven open system has been formulated by Deffner and Lutz $\cite{Deffner}$.
However, their time bound is derived from pure initial states and can not be
directly applied into the mixed initial states.
Motivated by the recent studies above, the authors in Ref. \cite{Sun}
employed an alternative fidelity definition different from relative entropy in order to
derive an easily computable QSL time bound for open systems whose initial states
can be chosen as either pure or mixed states and this QSL time is applicable to either Markovian or non-Markovian dynamics. Alongside these efforts, several attentions have been devoted to investigate the effects of entanglement on the QSL time in multiqubit systems (for instance, see \cite{Taddei, che}).

In this paper, we consider a bipartite system composed of two identical V-type three-level atoms as a two-qutrit system where each atom interacts with own independent reservoir. We use the QSL time bound obtained in $\cite{Sun}$ to evaluate the speed of evolution of the system under the effects of two-qtrit entanglement exhibited in the system. To this aim, we choose two local unitarily equivalent Werner states \cite{Horodecki3} and the Horodocki PPT (partial positive transpose) state \cite{Horodecki} which is not equivalent to the Werner states, as initial states, and obtained the QSL time for each of them in the respective entangled regions. When the degree of non-Markovianity in the system dynamics is considerable, surprisingly distinct behaviors for the QSL time are obtained for the equivalent Werner states in the entangled regions. In other words, under the condition of considerable non-Markovianity, effect of the same two-qutrit entanglement related to equivalent Werner states are completely different on the respective QSL times. It is interesting to note that for the non-equivalent Horodecki state, the calculated QSL time has an intermediate behavior relative to the cases obtained for the Werner states. We see that, although any local unitary operation does not increases the degree of entanglement (in the case of Werner states discussed in the present paper, the introduced local unitary operator does not change the degree of entanglement) but, as illustrated in the text, it is observed that the related QSL time changes under the mentioned local unitary operator effect.

The paper is organized as follows: In Sec. 2, we describe dynamics of a V-type three-level atomic system as an open system along with giving a quantitative description of non-Markovian behavior of that system. Also, we extend the dynamics of the V-type three-level system to a bipartite case. Sec. 3 is devoted to the obtaining of QSL time for the bipartite system and discussing about the connection between two-qutrit entanglement and the related QSL time. Finally, in Sec. 4, we present our conclusions.

\section{Dynamics of V-type three-level open systems}

In this section, we study at first, Markovian and non-Markovian dynamics of a V-type three level atom, as a qutrit, coupled to a dissipative environment. Each atom has two excited states each of them spontaneously decays into ground states such that the respective dipole moments of transitions may have interaction with each other.  We then extend the scheme into a two-qurit case which each of them coupled to independent reservoirs as depicted in Fig. 1. For a V-type atom, under the interaction with the same environment, the two upper levels $|2\rangle$ and $|1\rangle$ are coupled with the ground one, $|0\rangle$, with transition frequency $\omega_{2}$ and $\omega_{1}$ respectively.
The Hamiltonian of the total system is given by
\begin{eqnarray}
H=H_{S}+H_{E}+H_{I}=H_{0}+H_{I},
\end{eqnarray}
where
\begin{eqnarray}
H_{0}=\sum_{l=1}^{2}\omega_{l}\sigma_{+}^{l}\sigma_{-}^{l}+\sum_{k}\omega_{k}b_{k}^{\dagger}b_{k},
\end{eqnarray}
is the free Hamiltonian of the system and environment, and
\begin{eqnarray}
H_{I}=\sum_{l=l}^{2}\sum_{k}(g_{lk}\sigma_{+}^{l}b_{k}+g_{lk}^{\ast}\sigma_{-}^{l}b_{k}^{\dagger}),
\end{eqnarray}
is the interaction Hamiltonian. $\sigma_{\pm}^{l}$ ($l=1, 2$) are the raising and lowering operators of the levels $|2\rangle$ and $|1\rangle$ to the ground state $|0\rangle$. The index $k$ labels the different field modes of the environment with frequencies $\omega_{k}$, creation and annihilation operators $b_{k}^{\dagger}$, $b_{k}$ and coupling constants $g_{lk}$. In the interaction picture the total system obeys the Schr$\ddot{\mathrm{o}}$dinger equation
\begin{eqnarray}
i\frac{d}{dt}|\psi(t)\rangle=H_{int}|\psi(t)\rangle,
\end{eqnarray}
where
\begin{eqnarray}
H_{int}=\sum_{l=1}^{2}\sum_{k}(g_{lk}\sigma_{+}^{l}b_{k}e^{i(\omega_{l}-\omega_{k})t}+g_{lk}^{\ast}\sigma_{-}^{l}b_{k}^{\dagger}e^{-i(\omega_{l}-\omega_{k})t}),
\end{eqnarray}
and
\begin{eqnarray}
|\psi(t)\rangle=\sum_{l=0}^{2}c_{l}(t)|l\rangle_{S}\otimes|0\rangle_{E}+\sum_{k}c_{k}(t)|0\rangle_{S}\otimes|1_{k}\rangle_{E}.
\end{eqnarray}
Since $H_{int}|0\rangle_{S}\otimes|0\rangle_{E}=0$ then $c_{0}(t)$ will be invariant in time, while the amplitudes $c_{1}(t)$ and $c_{2}(t)$ will not. The time variations of these amplitudes is obtained by solving a system of differential equations which is easily derived from the Schr$\mathrm{\ddot{o}}$dinger equation
\begin{eqnarray}
\dot{c}_{l}(t)=-i\sum_{k}g_{lk}c_{k}(t)e^{i(\omega_{l}-\omega_{k})t},\quad l=1, 2,
\end{eqnarray}
\begin{eqnarray}
\dot{c}_{k}(t)=-i\sum_{l=1}^{2}g_{lk}^{\ast}c_{l}(t)e^{-i(\omega_{l}-\omega_{k})t}.
\end{eqnarray}
By assuming $c_{k}(0)=0$, i.e. there is no photon in the initial state, the solution of the second equation is inserted into the first one to get a closed equation for $c_{l}(t)$,
\begin{eqnarray}
\dot{c}_{l}(t)=-\sum_{m=1}^{2}\int_{0}^{t}f_{lm}(t-t')c_{m}(t')dt',\quad l=1, 2.
\end{eqnarray}
The kernel in Eqs. (9) can be expressed in term of spectral density $J(\omega)$ of the reservoir as follows
\begin{eqnarray}
f_{lm}(t-t')=\int_{0}^{t}d\omega J_{lm}(\omega)e^{i(\omega_{l}-\omega)t-i(\omega_{m}-\omega)t'}.
\end{eqnarray}
$J_{lm}(\omega)$ is chosen as Lorentzian distribution
\begin{eqnarray}
J_{lm}(\omega)=\frac{1}{2\pi}\frac{\gamma_{lm}\lambda^{2}}{(\omega_{0}-\Delta-\omega)^{2}+\lambda^{2}},
\end{eqnarray}
where $\Delta$ is the detuning of the atomic transition frequency
from the central frequency of the reservoir and $\lambda$ is the spectral
width of the coupling. $\gamma_{ii}=\gamma_{i}$ are the relaxation rates
of the two upper levels, and $\gamma_{ij}= \sqrt{\gamma_{i} \gamma_{j}} \theta$  $(i\neq j$ and $|\theta|\leq1)$ are responsible for the spontaneously generated interference (SGI) between the two decay channels
$|2\rangle \rightarrow |0\rangle$ and $|1\rangle \rightarrow |0\rangle$. $\theta$ depends on
the relative angle between two dipole moment elements related to the mentioned transitions. $\theta=0$ means that the dipole moments of two transitions are perpendicular to each other and this is corresponding to the case that there is no SGI between two decay channels. On the other hand, $\theta=\pm1$ indicating that the two dipole moment are parallel or antiparallel, corresponding to the strongest SGI between two decay channels.

If we take Laplace transform from Eqs. (9), it becomes
\begin{eqnarray}
\left(
  \begin{array}{c}
    p\tilde{c}_{2}(p)-c_{2}(0) \\
    p\tilde{c}_{1}(p)-c_{1}(0) \\
  \end{array}
\right)=-\frac{\lambda}{2(p+\lambda)}\left(
                                       \begin{array}{cc}
                                         \gamma_{2} & -\sqrt{\gamma_{1}\gamma_{2}}\theta \\
                                         \sqrt{\gamma_{1}\gamma_{2}}\theta & \gamma_{1} \\
                                       \end{array}
                                     \right)\left(
                                                    \begin{array}{c}
                                                      \tilde{c}_{2}(p) \\
                                                      \tilde{c}_{1}(p) \\
                                                    \end{array}
                                                  \right),
\end{eqnarray} where $\tilde{c}_{l}(p)=\mathcal{L}[c_{l}(t)]=\int_{0}^{\infty} c_{l}(t) e^{-pt} dt $ ($l=1, 2$), is the Laplace transform of $c_{l}(t)$. By considering the subspace spanned by $\{|2\rangle, |1\rangle\}$, and doing the following unitary transformation
\begin{eqnarray}
\mathcal{U}=\left(
  \begin{array}{cc}
    \sqrt{\frac{q+\gamma_{1}-\gamma_{2}}{2q}} & -\sqrt{\frac{q-\gamma_{1}+\gamma_{2}}{2q}} \\
    \sqrt{\frac{q-\gamma_{1}+\gamma_{2}}{2q}} & \sqrt{\frac{q+\gamma_{1}-\gamma_{2}}{2q}} \\
  \end{array}
\right),
\end{eqnarray}
on the Eqs. (12) and taking inverse Laplace transform we get
\begin{eqnarray}
c_{\pm}(t)=G_{\pm}(t)c_{\pm}(0),
\end{eqnarray}
where $c_{\pm}(0)=\frac{1}{\sqrt{2q}}\left(
                                       \begin{array}{c}
                                         c_{2}(0)\sqrt{q\pm\gamma_{1}\mp\gamma_{2}}\mp c_{1}(0)\sqrt{q\mp\gamma_{1}\pm\gamma_{2}} \\
                                       \end{array}
                                     \right)
$. In Eqs. (14),
\begin{eqnarray}
G_{\pm}(t)=e^{-\lambda t/2}\left(
                                    \begin{array}{c}
                                      \mathrm{cosh}(\frac{d_{\pm} t}{2})+\frac{\lambda}{d_{\pm}}\mathrm{sinh}(\frac{d_{\pm} t}{2}) \\
                                    \end{array}
                                  \right),
\end{eqnarray}
with $d_{\pm}=\sqrt{\lambda^{2}-2\lambda \gamma_{\pm}}$, $\gamma_{\pm}=\frac{\gamma_{1}+\gamma_{2}\pm q}{2}$ and $q=\sqrt{(\gamma_{1}-\gamma_{2})^{2}+4\gamma_{1}\gamma_{2}\theta^{2}}$. It should be noted that the extended unitary transformation on the whole space of the system spanned by $\{|2\rangle, |1\rangle, |0\rangle\}$, becomes as
\begin{eqnarray}
U=\left(
  \begin{array}{ccc}
    \sqrt{\frac{q+\gamma_{1}-\gamma_{2}}{2q}} & -\sqrt{\frac{q-\gamma_{1}+\gamma_{2}}{2q}} & 0\\
    \sqrt{\frac{q-\gamma_{1}+\gamma_{2}}{2q}} & \sqrt{\frac{q+\gamma_{1}-\gamma_{2}}{2q}} & 0\\
    0 & 0 & 1\\
  \end{array}
\right).
\end{eqnarray}

By these Considerations, the density matrix of a V-type system at time $t$ becomes
\begin{eqnarray}
\varrho_{S}(t)=\left(
                \begin{array}{ccc}
                  |G_{+}(t)|^{2}|c_{+}(0)|^{2} & G_{+}(t)c_{+}(0)G^{\ast}_{-}(t)c^{\ast}_{-}(0) & G_{+}(t)c_{+}(0)c_{0}^{\ast} \\\\
                  G_{+}^{\ast}(t)c_{+}(0)^{\ast}G_{-}(t)c_{-}(0) & |G_{-}(t)|^{2}|c_{-}(0)|^{2} & G_{-}(t)c_{-}(0)c_{0}^{\ast} \\\\
                  G^{\ast}_{+}(t)c^{\ast}_{+}(0)c_{0} & G^{\ast}_{-}(t)c^{\ast}_{-}(0)c_{0} &\begin{array}{c}
                                                                                               1-|G_{+}(t)|^{2}|c_{+}(0)|^{2} \\
                                                                                               -|G_{-}(t)|^{2}|c_{-}(0)|^{2}
                                                                                             \end{array}
                   \\
                \end{array}
              \right).
\end{eqnarray}
Clearly, the density matrix in Eqs. (17) is, indeed, the time development of the state $\varrho_{S}(0)=U|\psi(0)\rangle\langle\psi(0)|U^{\dagger}$ in which $|\psi(0)\rangle$ is the state (6) at time $t=0$. It is very instructive to write the density matrix (17) in the Krauss representation. It can be easily obtained that
\begin{eqnarray}
\varrho_{S}(t)=\sum_{i=1}^{3}\mathcal{K}_{i}\varrho_{S}(0)\mathcal{K}^{\dagger}_{i},
\end{eqnarray}
with $\sum_{i=1}^{3}\mathcal{K}^{\dagger}_{i}\mathcal{K}_{i}=I_{3}$, where $I_{3}$ is identity operator for the Hilbert space of a three-level system and
\begin{eqnarray}
\begin{array}{c}
  \mathcal{K}_{1}=\left(
                    \begin{array}{ccc}
                      G_{+}(t) & 0 & 0 \\
                      0 & G_{-}(t) & 0 \\
                      0 & 0 & 1 \\
                    \end{array}
                  \right),
   \\\\
  \mathcal{K}_{2}=\left(
                    \begin{array}{ccc}
                      0 & 0 & 0 \\
                      0 & 0 & 0 \\
                      \sqrt{1-|G_{+}(t)|^{2}} & 0 & 0 \\
                    \end{array}
                  \right),
   \\\\
  \mathcal{K}_{3}=\left(
                    \begin{array}{ccc}
                      0 & 0 & 0 \\
                      0 & 0 & 0 \\
                      0 & \sqrt{1-|G_{-}(t)|^{2}} & 0 \\
                    \end{array}
                  \right).
\end{array}
\end{eqnarray}
Also, the original density matrix is obtained as $\rho_{S}(t)=U^{\dagger}\varrho_{S}(t)U$ and the respective Krauss operators are $K_{i}=U^{\dagger}\mathcal{K}_{i}U$ ($i=1, 2, 3$), with explicit forms
\begin{eqnarray}
K_{1}=\left(
        \begin{array}{ccc}
          G_{+}\frac{q+\gamma_{1}-\gamma_{2}}{2q}+G_{-}\frac{q-\gamma_{1}+\gamma_{2}}{2q} & (-G_{+}+G_{-})\frac{\sqrt{q^{2}-(\gamma_{1}-\gamma_{2})^2}}{2q} & 0 \\\\
          (-G_{+}+G_{-})\frac{\sqrt{q^{2}-(\gamma_{1}-\gamma_{2})^2}}{2q} & G_{+}\frac{q-\gamma_{1}+\gamma_{2}}{2q}+G_{-}\frac{q+\gamma_{1}-\gamma_{2}}{2q} & 0 \\\\
          0 & 0 & 1 \\
        \end{array}
      \right),
\end{eqnarray}
\begin{eqnarray}
K_{2}=\frac{\sqrt{(1-|G_{+}|^2)}}{2q}\left(
        \begin{array}{ccc}
          0 & 0 & 0 \\\\
          0 & 0 & 0 \\\\
          \sqrt{q+\gamma_{1}-\gamma_{2}} & -\sqrt{q-\gamma_{1}+\gamma_{2}} & 0 \\
        \end{array}
      \right)
,
\end{eqnarray}
\begin{eqnarray}
K_{3}=\frac{\sqrt{(1-|G_{-}|^2)}}{2q}\left(
        \begin{array}{ccc}
          0 & 0 & 0 \\\\
          0 & 0 & 0 \\\\
          \sqrt{q-\gamma_{1}+\gamma_{2}} & \sqrt{q+\gamma_{1}-\gamma_{2}} & 0 \\
        \end{array}
      \right)
,
\end{eqnarray}
hence, it is concluded that
$\sum_{i=1}^{3}K^{\dagger}_{i}K_{i}=I_{3}$ and $\rho_{S}(t)=\sum_{i=1}^{3}K_{i}\rho_{S}(0)K^{\dagger}_{i}$.
It should be noted that the initial state of a V-type atom, i.e. $\rho_{S}(0)$, can be generally considered as a mixed state so its time development can also be obtained easily.

In this paper, we consider the case in which the two upper atomic levels are degenerated and the atomic transitions are in resonant with the central frequency of the reservoir, i.e. $\omega_{1}=\omega_{2}=\omega_{0}$ and $\Delta=0$. Under this consideration, we assume that the relaxation rates of two decay channels are equal, i.e. $\gamma_{1}=\gamma_{2}=\gamma$. Therefore, the statement in Eqs. (15), takes a simple form. In order to study the non-Markovian dynamics of the V-type atom, exploiting a non-Markovian measure is inevitable. In Ref. $\cite{Breuer}$, Breuer $et. al$ introduced a measure of non-Markovianity, which is based on
the reverse flow of information from the reservoir back to the system as follows
\begin{eqnarray}
  \mathcal{N}=max_{\rho_{1,2}(0)} \int_{\eta>0} \eta[t,\rho_{1,2}(0)]dt,
\end{eqnarray}
where
\begin{eqnarray}
  \eta[t,\rho_{1,2}(0)]=\frac{d}{dt} D[\rho_{1}(t),\rho_{2}(t)],
\end{eqnarray}
indicates the changing rate of the trace distance of a pair
of states denoted by $D[\rho_{1}(t),\rho_{2}(t)]=\frac{1}{2} Tr|\rho_{1}(t)-\rho_{2}(t)|$ with the trace norm definition for an operator such as $A$ as $|A|=\sqrt{A^{\dag}A}$. In general, in order to calculate the $\mathcal{N}$ in Eqs. (23), it is required to cover any pair of initial states. However, in Ref. $\cite{Wen}$ the optimized pair of initial states for a V-type three-level atom is obtained analytically.
Fig. 2, shows the non-Markovian behavior of the system in terms of $\gamma$, $\lambda$ and $\theta$. When $\theta$ ($|\theta|$) increases the SGI becomes more and more strong and this leads to the improvement of the non-Markovian behavior. Similar process takes place by increasing the system environment coupling $\gamma$. But the non-Markovian measure decreases in terms of the spectral width of the coupling $\lambda$.

Extending above method for dynamics of a system consists of two identical V-type atoms each of them independently interact with a Lorentzian environment is a trivial task. Let's $\rho_{S}(0)$ be defined as a density matrix of a two-qutrit V-type system on the $3\otimes3$ Hilbert space. So in this way, its time development at time $t$ becomes
\begin{eqnarray}
\rho_{S}(t)=\sum_{k,l=1}^{3}K_{k,l}\rho_{S}(0)K^{\dagger}_{k,l}, \quad \sum_{k,l=1}^{3}K^{\dagger}_{k,l}K_{k,l}=I_{3}\otimes I_{3},
\end{eqnarray}
where $K_{k,l}=K_{k}\otimes K_{l}$. In the next section, we consider possible types of entanglement for the two-qutrit V-type system at initial time, and investigate their effects on the related QSL time throughout various non-Markovian dynamics.

\section{Quantum speed limit time for a two-qutrit V-type system}

In this section, we introduce an easily computable QSL time for a two-qutrit V-type open system described in the previous section, whose initial states can be chosen as either pure or mixed. To this aim, we employ an alternative fidelity definition, as a distance measure of two quantum states introduced in \cite{Wa} as
\begin{eqnarray}
F(\rho_{0},\rho_{t})=\frac{Tr(\rho_{0},\rho_{t})}{\sqrt{Tr(\rho_{0}^{2})Tr(\rho_{t}^{2})}},
\end{eqnarray}
to calculate the QSL time bound, as derived in \cite{Sun} (note that $\rho(t)\equiv\rho_{t}$). According to the Ref. \cite{Sun}, the derived QSL time bound which is applicable to either Markovian and
non-Markovian dynamics, is as the following form
\begin{eqnarray}
\tau\geq\tau_{QSL}=\frac{|1-F(\rho_{0},\rho_{\tau})|}{X(\tau)},
\end{eqnarray}
where
\begin{eqnarray}
X(\tau)=\frac{2}{\tau}\int_{0}^{\tau}\sqrt{\frac{Tr(\dot{\rho}_{t}^{2})}{Tr(\rho_{t}^{2})}}dt,
\end{eqnarray}
by denoting that $\dot{\rho}_{t}$ is the time derivative of the state $\rho_{t}$ and $\tau$ is the actual driving time.

At the first step, we consider the initial state of the two-qutrit system to be the following Werner state $\cite{Horodecki3}$
\begin{eqnarray}
\begin{array}{c}
  \rho^{p}_{0}(0)=(1-p) \frac{I_{3}\otimes I_{3}}{9}+p|\psi_{0} \rangle \langle \psi_{0}|,
\end{array}
\end{eqnarray}
where $p\in[0,1]$ and $|\psi_{0}\rangle=\frac{1}{\sqrt{3}}(|00\rangle+|11\rangle+|22\rangle)$.
This state for $p\leq 1/4$, is separable whereas for $p>1/4$ entangled. In fact, since it violate reduction criterion of separability so such Werner state are distillable $\cite{Horodecki2}$.

By this consideration, in Fig. 3, Fig. 4 and Fig. 5, we have shown the $\tau_{QSL}$ (QSL time) in term of $\gamma$ for parameter values $\lambda=0.1$, $\lambda=1$ and $\lambda=10$ (for fixed $\theta=1$) respectively. In Fig. 3 for $\lambda=0.1$, as the non-Markovianity of the system grows up in term of $\gamma$, $\tau_{QSL}$ suddenly decreases. It is observed that the decrement rate of $\tau_{QSL}$ is proportional to the degree of entanglement of Werner state. In other words, more entanglement leads to evolution with more speed. In Fig. 4 for $\lambda=1$, $\tau_{QSL}$ fluctuationally decreases and the fluctuations amplitude is proportional to the degree of entanglement. Fig. 5, for $\lambda=10$, shows a smooth decreasing of $\tau_{QSL}$ in term of $\gamma$ and the speed of evolution of the system is always inversely proportional to the degree of entanglement.

If we perform the following local unitary shift operator as
\begin{eqnarray}
  \mathcal{U}=\left(
                \begin{array}{ccc}
                  1 & 0 & 0 \\
                  0 & 1 & 0 \\
                  0 & 0 & 1 \\
                \end{array}
              \right)\otimes\left(
                              \begin{array}{ccc}
                                0 & 1 & 0 \\
                                0 & 0 & 1 \\
                                1 & 0 & 0 \\
                              \end{array}
\right),
\end{eqnarray}
on the state (29), we obtain the other Werner state as our another initial state as follows
\begin{eqnarray}
\begin{array}{c}
  \rho^{p}_{1}(0)=(1-p) \frac{I_{3}\otimes I_{3}}{9}+p|\psi_{1} \rangle \langle \psi_{1}|,
\end{array}
\end{eqnarray}
where $|\psi_{1}\rangle=\frac{1}{\sqrt{3}}(|01\rangle+|12\rangle+|20\rangle)$. Since the local shift operator (30) does not change the amount of entanglement of the state (29) but we obtain a surprisingly different result for the QSL time when the initial state is chosen as (31). As observed from Fig. 6, QSL time for $\lambda=0.1$ does not generally suffer sudden decrement in comparison to the previous one as shown in Fig. 3. In other words, in contrast to the result of Fig. 3, the decrement rate of $\tau_{QSL}$ is inversely proportional to the degree of entanglement of the Werner state in (31). Consequently, more entanglement leads to evolution with less speed. In Fig. 7 for $\lambda=1$, $\tau_{QSL}$ fluctuationally decreases with less amplitude in comparison to the Fig. 4 and the fluctuations amplitude is also proportional to the degree of entanglement. Fig. 8, for $\lambda=10$, shows that the effect of two-qutrit entanglement on the QSL time for the initial state (31) is is similar to the case presented in Fig. 5. It is concluded that as the degree of non-Markovianity is considerable in the system (see Fig. 2), the effects of two-qutrit entanglement of two local unitarily equivalent Werner states on the corresponding QSL times will be completely different.

In the next step, we take the well-known Horodecki state $\cite{Horodecki}$ as an another initial state for the two-qutrit V-type system as follows
\begin{eqnarray}
  \rho^{\alpha}(0)=\frac{2}{7}|\psi_{0} \rangle \langle \psi_{0}|+\frac{\alpha}{7}\sigma_{+}+\frac{5-\alpha}{7}\sigma_{-},
\end{eqnarray}
where
\begin{eqnarray}
\begin{array}{c}
  \sigma_{+}=\frac{1}{3}(|01\rangle \langle01|+|12\rangle \langle12|+|20\rangle \langle20|),\\\\
  \sigma_{-}=\frac{1}{3}(|10\rangle \langle10|+|21\rangle \langle21|+|02\rangle \langle02|),
\end{array}
\end{eqnarray}
with $0\leq \alpha \leq5$. $\rho^{\alpha}(0)$ is free entangled for $0\leq \alpha <1$, bound entangled for $1\leq \alpha <2$ and is separable
for $2\leq \alpha \leq3$. Notice that $\rho^{\alpha}$ is
invariant under the simultaneous change of $\alpha \longrightarrow 5-\alpha$
and interchange of the parties so by these operations, $\rho^{\alpha}$ is free entangled for $4< \alpha \leq5$, bound entangled for $3< \alpha \leq4$ and is separable for $2\leq \alpha \leq3$. It was shown in \cite{Nag} that the entanglement of $\rho^{\alpha}$ is decreased, in term of parameter $\alpha$, in the interval $\alpha\in[0, 2]$ and increased in the interval $\alpha\in[3, 5]$ symmetrically. Therefore, we only analysis effect of entanglement of $\rho^{\alpha}(0)$ on the related QSL for $\alpha\in[0, 2]$.

In similar way discussed above, for the Horodecki state (32), we have shown the behaviors of $\tau_{QSL}$ in term of $\gamma$ for parameter values $\lambda=0.1$, $\lambda=1$ and $\lambda=10$ (for fixed $\theta=1$) throughout the Fig. 9, Fig. 10 and Fig. 11 respectively. It is evident that the Horodecki state (32) is completely different from the previous Werner states such that we can not obtain it from the Werner states by any local unitary operations. However, we observe that the effect of entanglement (free and PPT entanglement) of Horodecki state on the related QSL time has an intermediate behavior in comparison to the effect of entanglement of Werner states (29) and (31).

Anther point which should be noted in this paper is that though two Werner states (29) and (31) are local unitarily equivalent but the QSL time of (31) is always greater than the case of (29) (compare Fig. 3, Fig. 4 and Fig. 5 with Fig. 6, Fig. 7 and Fig. 8 respectively). In other words, performing the local unitary operation (30) on the Werner state (29) leads to reduce the speed of evolution of the system. Also, the QSL time for the non-equivalent Horodecki state (32) relies between the QSL times obtained from the Werner states.

At the end, by interchanging the atoms (atom1 $\rightleftharpoons$ atom2) in (31), another Werner state is obtained. By considering this state as an initial state, the obtained results are the same as the results of the state (31).

\section{Conclusions}
In summary, we investigated effect of two-qutrit entanglement on the QSL time of a bipartite V-type atomic open system. It was shown that as the non-Markovianity in the system dynamics is considerable, the entanglement effect of two local unitarily equivalent Werner states are completely different. Also, we found out that the entanglement effect of non-equivalent Horodecki state on the related QSL time has an intermediate behavior relative to the previous cases. We observed that although the local unitary shift operator (30) does not change the two-qutrit entanglement but, in turn, may affect speed of evolution of the system considerably. Generally speaking, this point encourage us to examine the effect of other local unitary operators (local quantum gates) on the speed of evolution of systems in the future studies, which is of essential importance in quantum information processing.

\newpage
Fig. 1. A system of two identical V-type three-level atoms each of them interacts with own independent reservoir.

\begin{figure}
\centering
\includegraphics[width=445 pt]{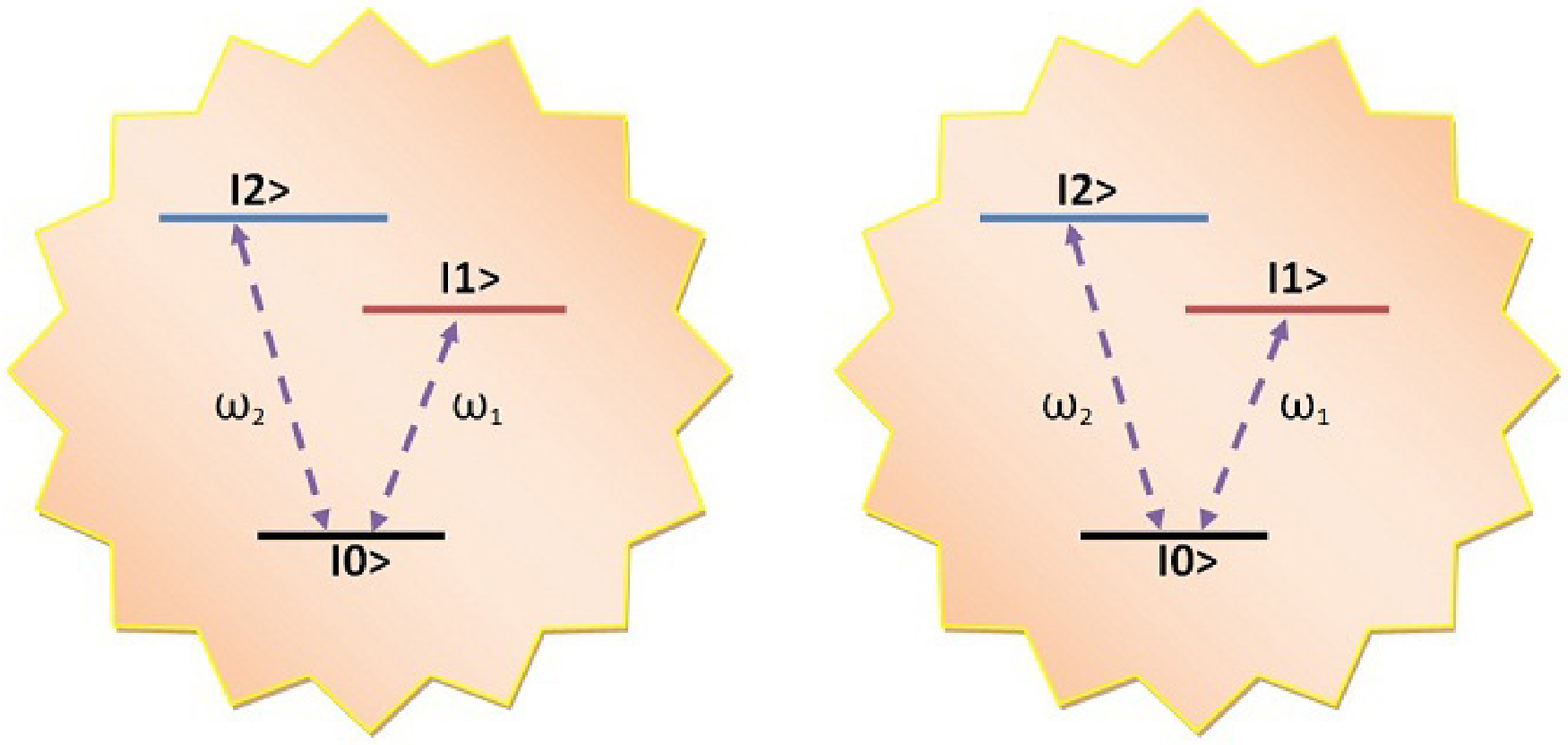}
\caption{} \label{Fig1}
\end{figure}

\newpage
Fig. 2. Measure of non-Markovianity $\mathcal{N}$ versus $\gamma$ with $\theta=1$ (solid line) and $\theta=0.6$ (dashed line).
\begin{figure}
\centering
\includegraphics[width=445 pt]{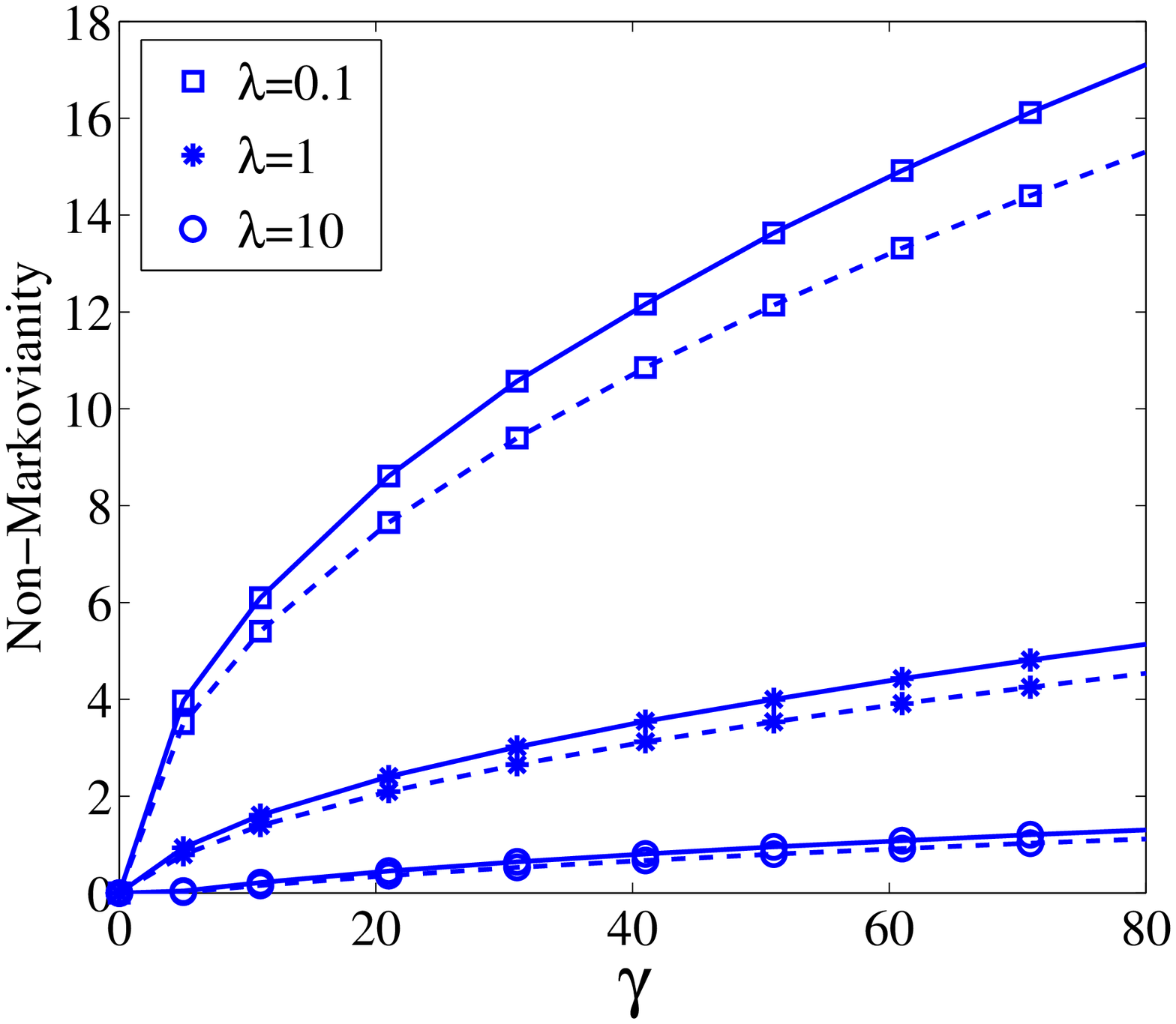}
\caption{} \label{Fig1}
\end{figure}

\newpage
Fig. 3. QSL time versus $\gamma$ with $\theta=1$ and $\lambda=0.1$. The Werner state in (29) is the initial state where $p$ is chosen in the entangled region. As the non-Markovianity of the system grows up in term of $\gamma$, the decrement rate of QSL time is proportional to the entanglement. The actual driving time is $\tau=1$.
\begin{figure}
\centering
\includegraphics[width=445 pt]{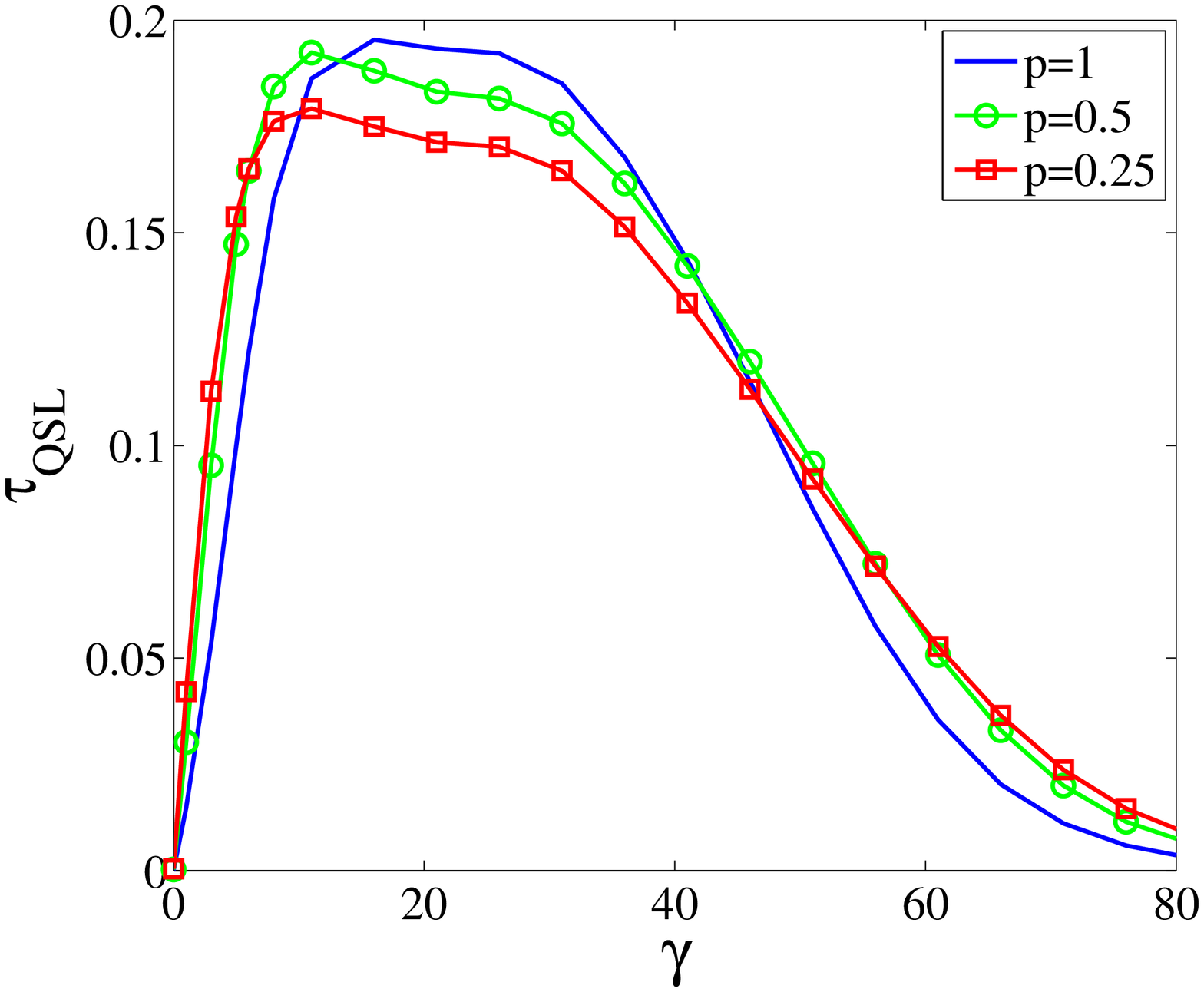}
\caption{} \label{Fig1}
\end{figure}

\newpage
Fig. 4. QSL time versus $\gamma$ with $\theta=1$ and $\lambda=1$. The Werner state in (29) is the initial state where $p$ is chosen in the entangled region. The fluctuation of QSL time is proportional to the entanglement.
\begin{figure}
\centering
\includegraphics[width=445 pt]{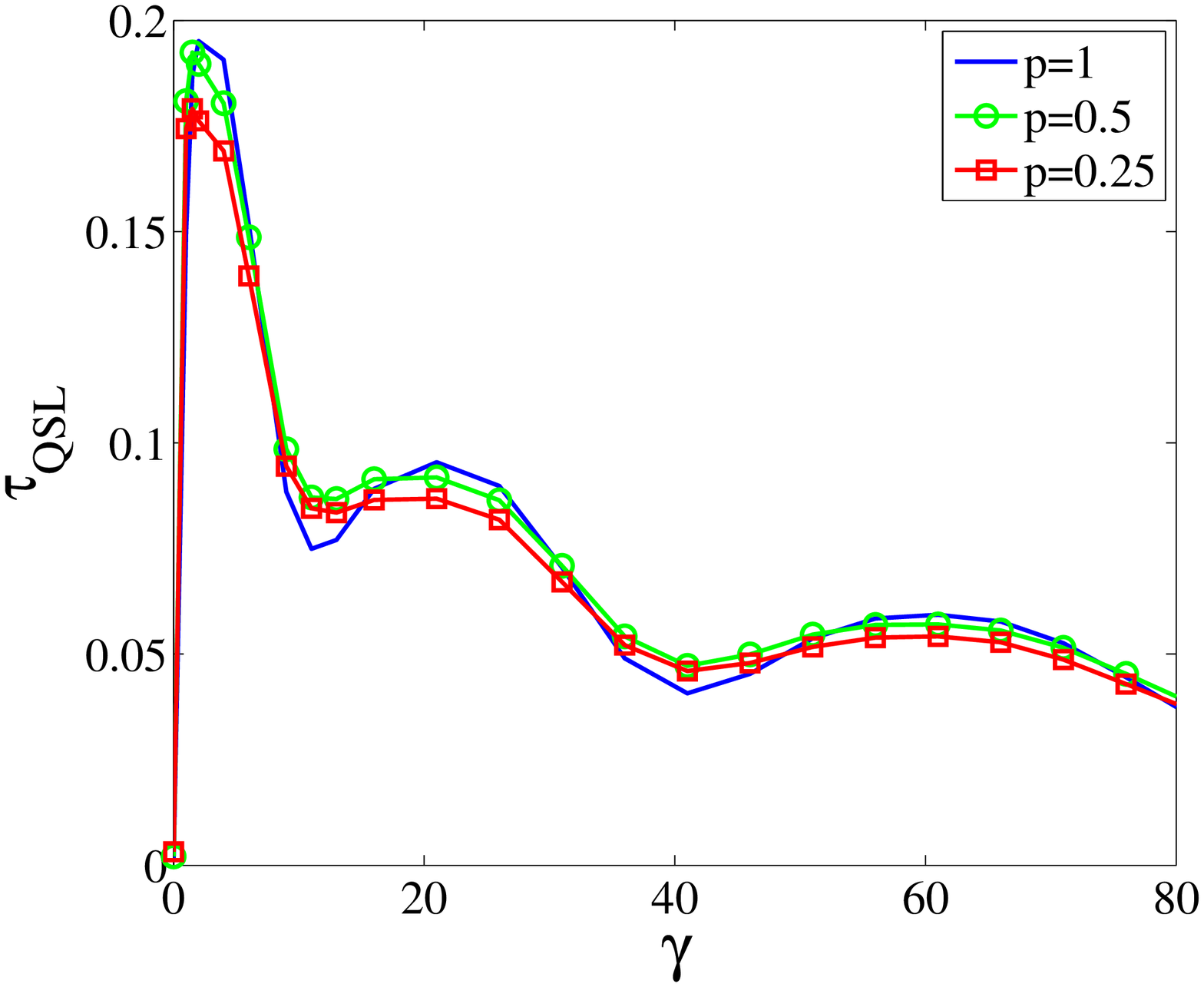}
\caption{} \label{Fig1}
\end{figure}

\newpage
Fig. 5. QSL time versus $\gamma$ with $\theta=1$ and $\lambda=10$. The Werner state in (29) is the initial state where $p$ is chosen in the entangled region. Speed of evolution of the system is inversely proportional to the degree of entanglement.
\begin{figure}
\centering
\includegraphics[width=445 pt]{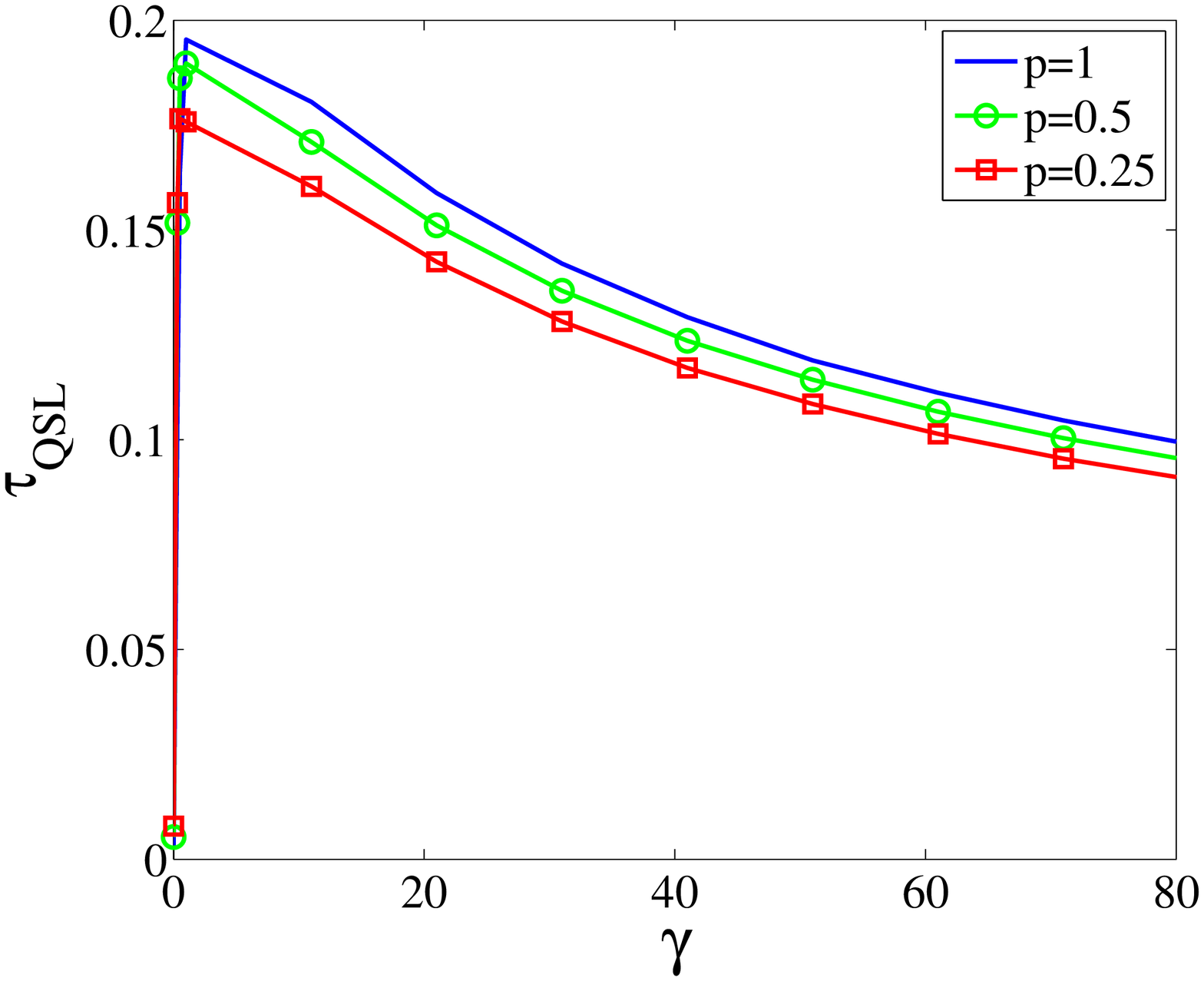}
\caption{} \label{Fig1}
\end{figure}

\newpage
Fig. 6. QSL time versus $\gamma$ with $\theta=1$ and $\lambda=0.1$. The Werner state in (31) is the initial state where $p$ is chosen in the entangled region. In contrast to the case presented in Fig. 3, as the non-Markovianity of the system grows up in term of $\gamma$, the decrement rate of QSL time is inversely proportional to the entanglement. Also, the QSL time is always greater than the corresponding case of Fig. 3 in the entangled region.
\begin{figure}
\centering
\includegraphics[width=445 pt]{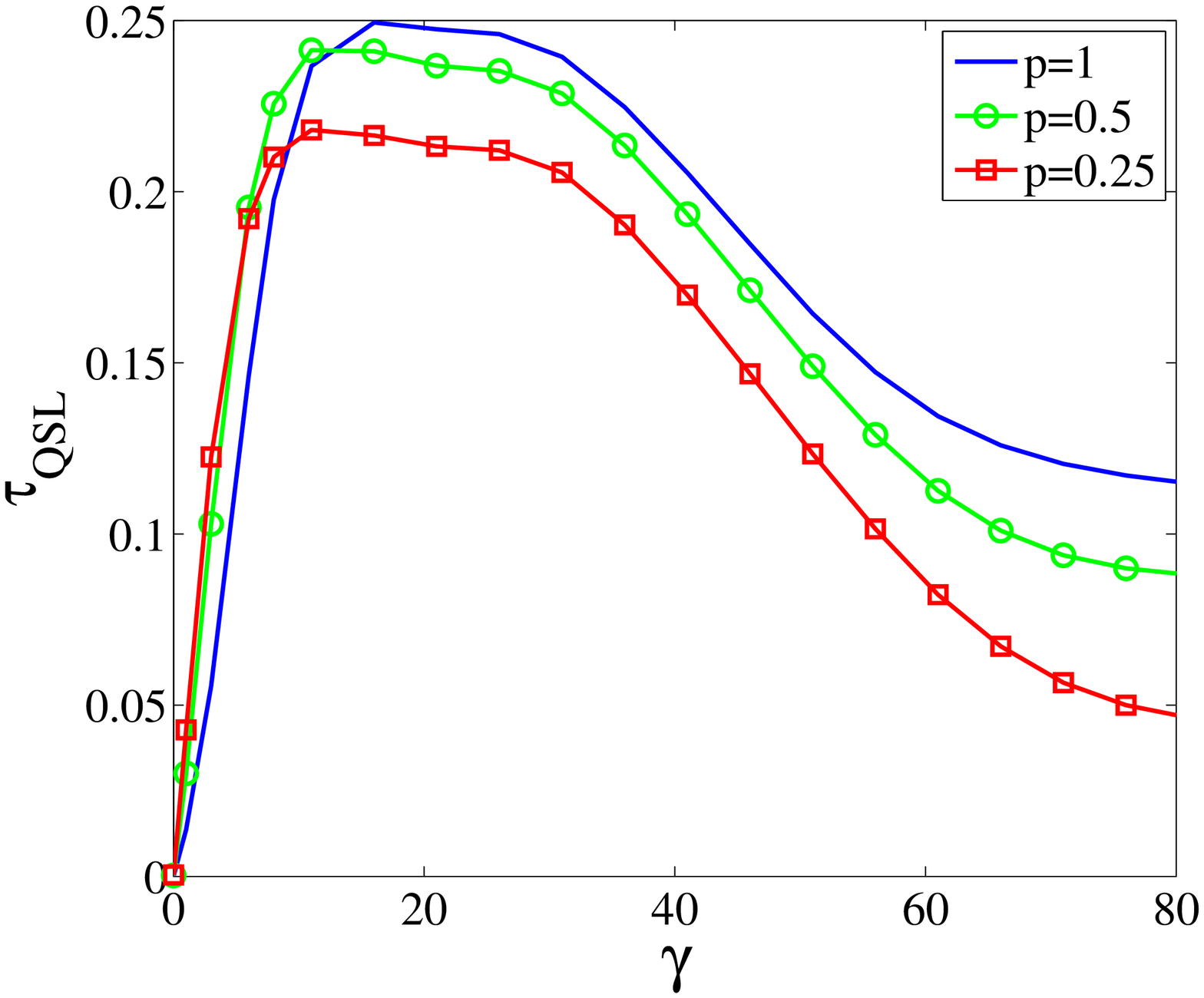}
\caption{} \label{Fig1}
\end{figure}

\newpage
Fig. 7. QSL time versus $\gamma$ with  $\theta=1$ and $\lambda=1$. The Werner state in (31) is the initial state where $p$ is chosen in the entangled region. QSL time fluctuationally decreases with less amplitude than the respective case brought in Fig. 4 and the fluctuations amplitude is also proportional to the degree of entanglement. The QSL time for this case is also greater than the corresponding case of Fig. 4 in the entangled region.
\begin{figure}
\centering
\includegraphics[width=445 pt]{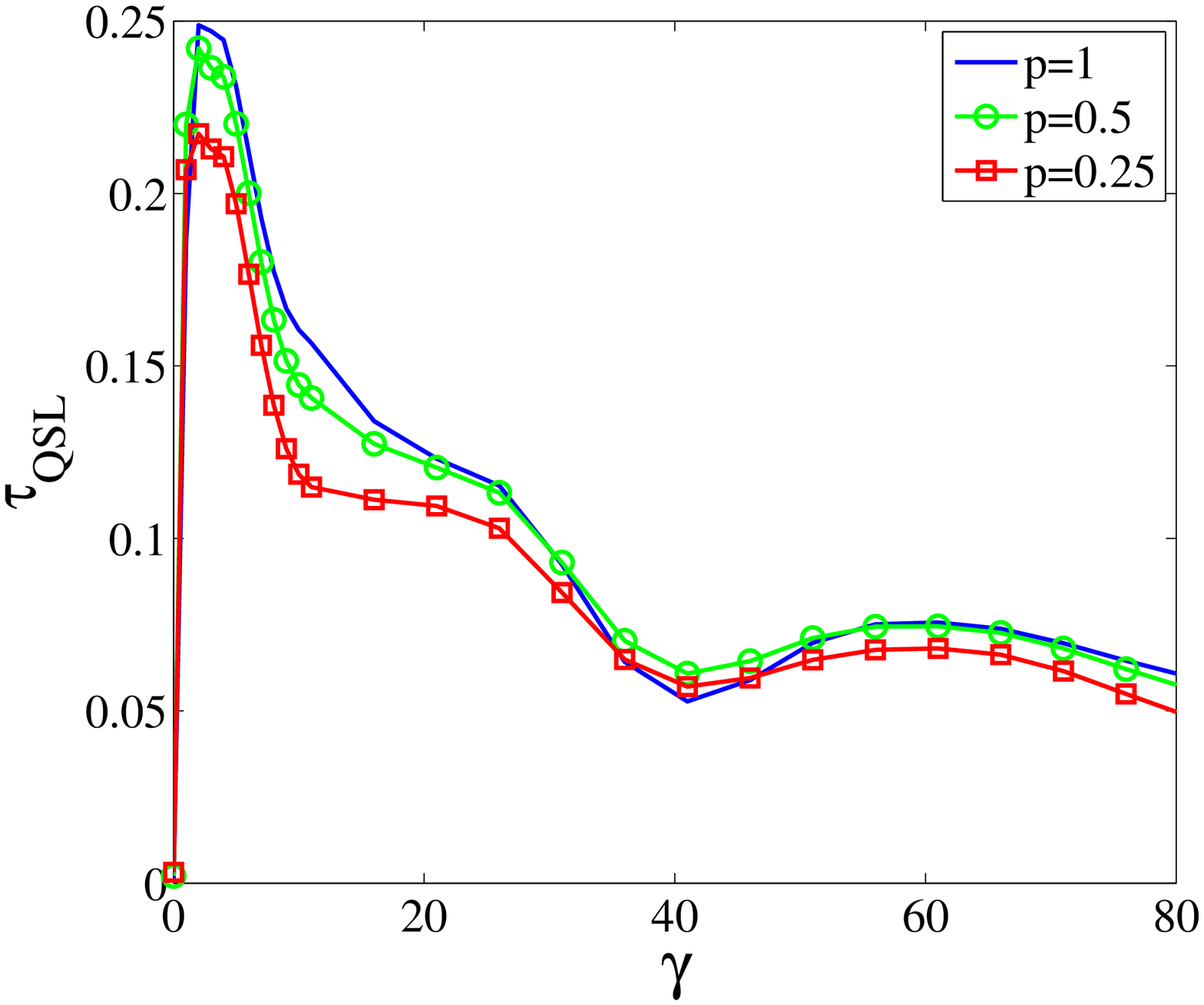}
\caption{} \label{Fig1}
\end{figure}

\newpage
Fig. 8. QSL time versus $\gamma$ with $\theta=1$ and $\lambda=10$.
The Werner state in (31) is the initial state where $p$ is chosen in the entangled region. Effect of entanglement on the QSL time is similar to the case presented in Fig. 5. Also, the QSL time for this case is greater than the corresponding case of Fig. 5 in the entangled region.
\begin{figure}
\centering
\includegraphics[width=445 pt]{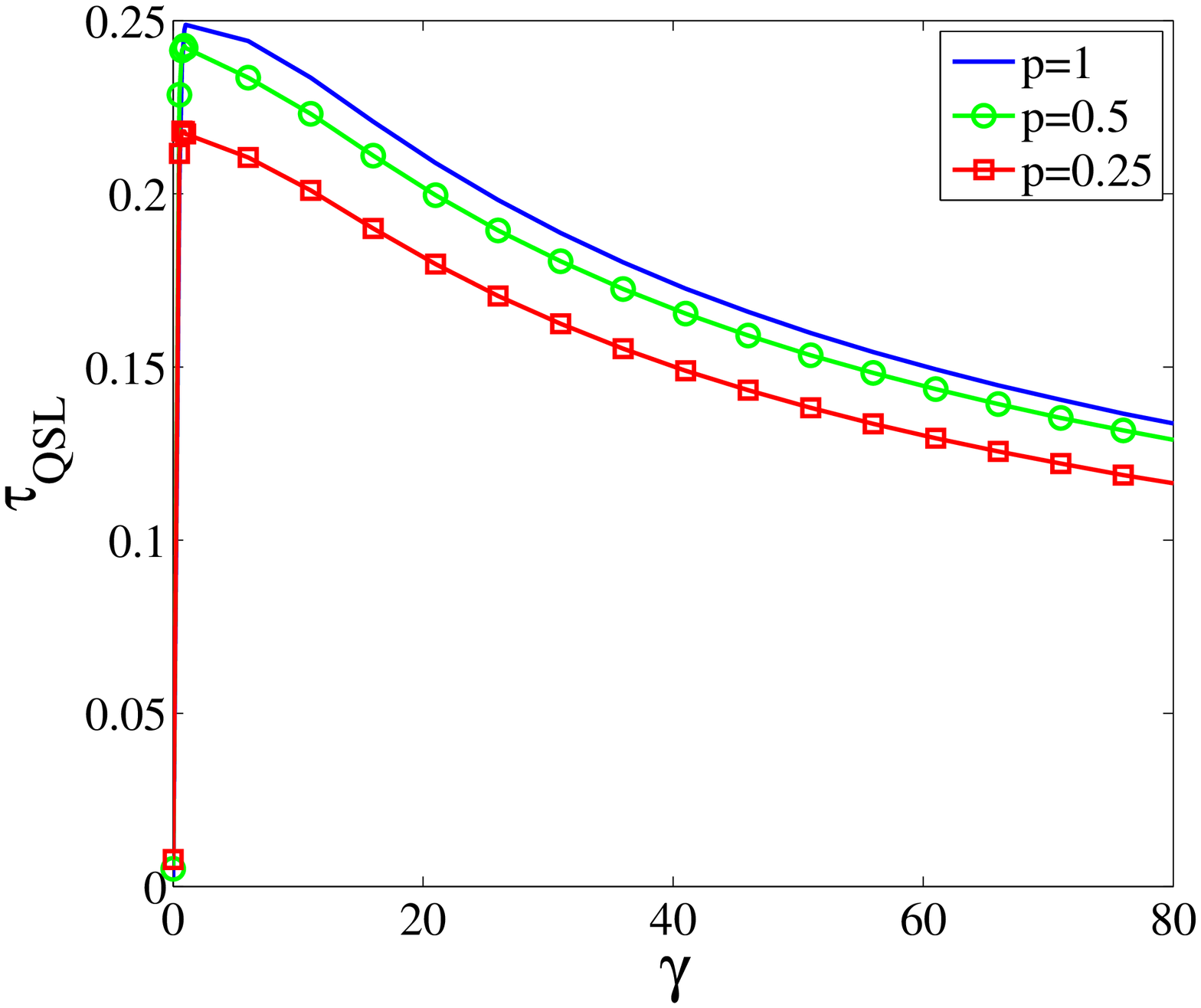}
\caption{} \label{Fig1}
\end{figure}

\newpage
Fig. 9. QSL time versus $\gamma$ with $\theta=1$ and $\lambda=0.1$.
The Horodecki state in (32) is the initial state where $\alpha$ is chosen in the entangled region (free and PPT entangled region). The QSL time for this state relies between the corresponding cases of Werner states presented in Fig. 3 and Fig. 6.
\begin{figure}
\centering
\includegraphics[width=445 pt]{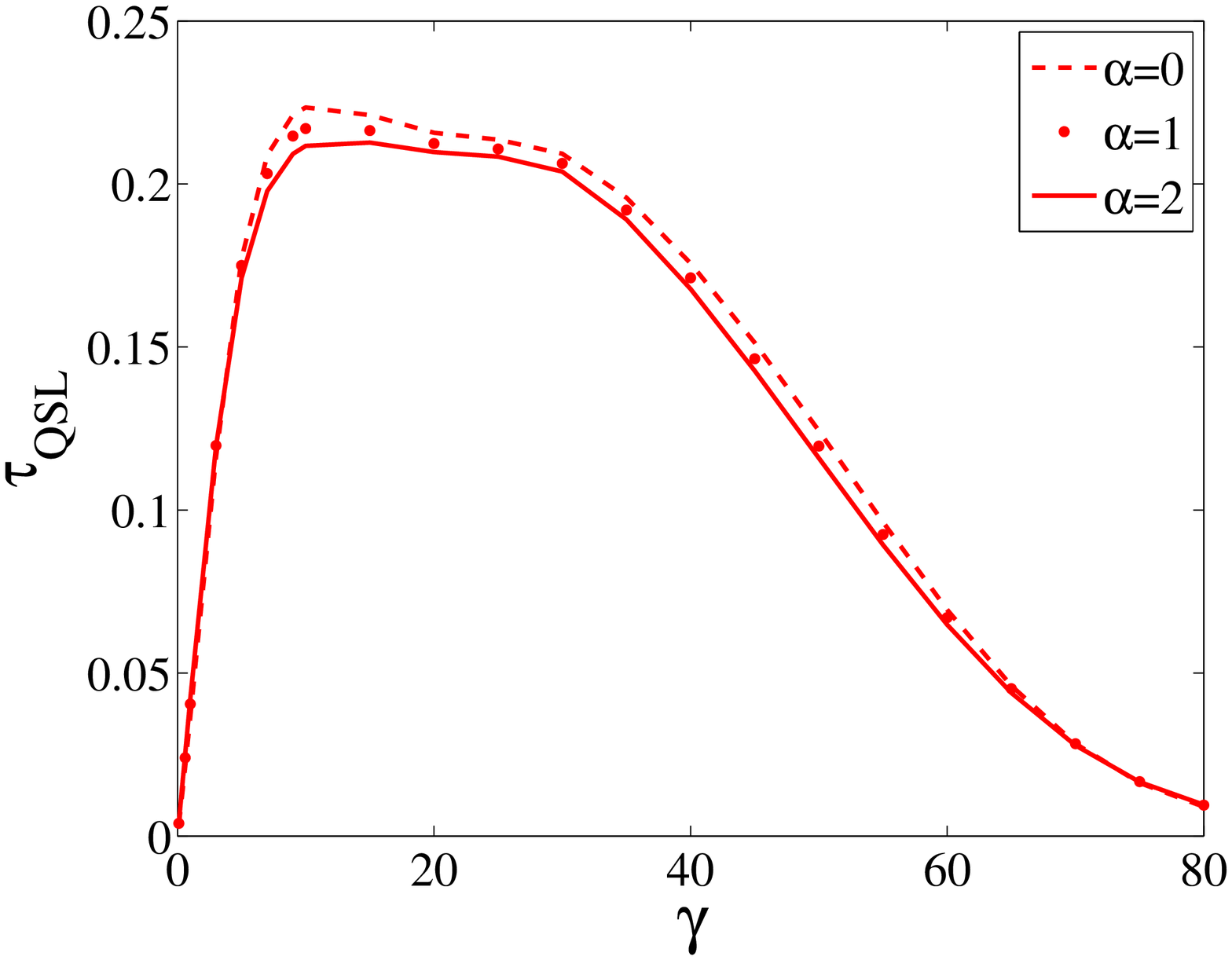}
\caption{} \label{Fig1}
\end{figure}

\newpage
Fig. 10. QSL time versus $\gamma$ with $\theta=1$ and $\lambda=1$. The Horodecki state in (32) is the initial state where $\alpha$ is chosen in the entangled region (free and PPT entangled region). The QSL time for this state relies between the corresponding cases of Werner states presented in Fig. 4 and Fig. 7.
\begin{figure}
\centering
\includegraphics[width=445 pt]{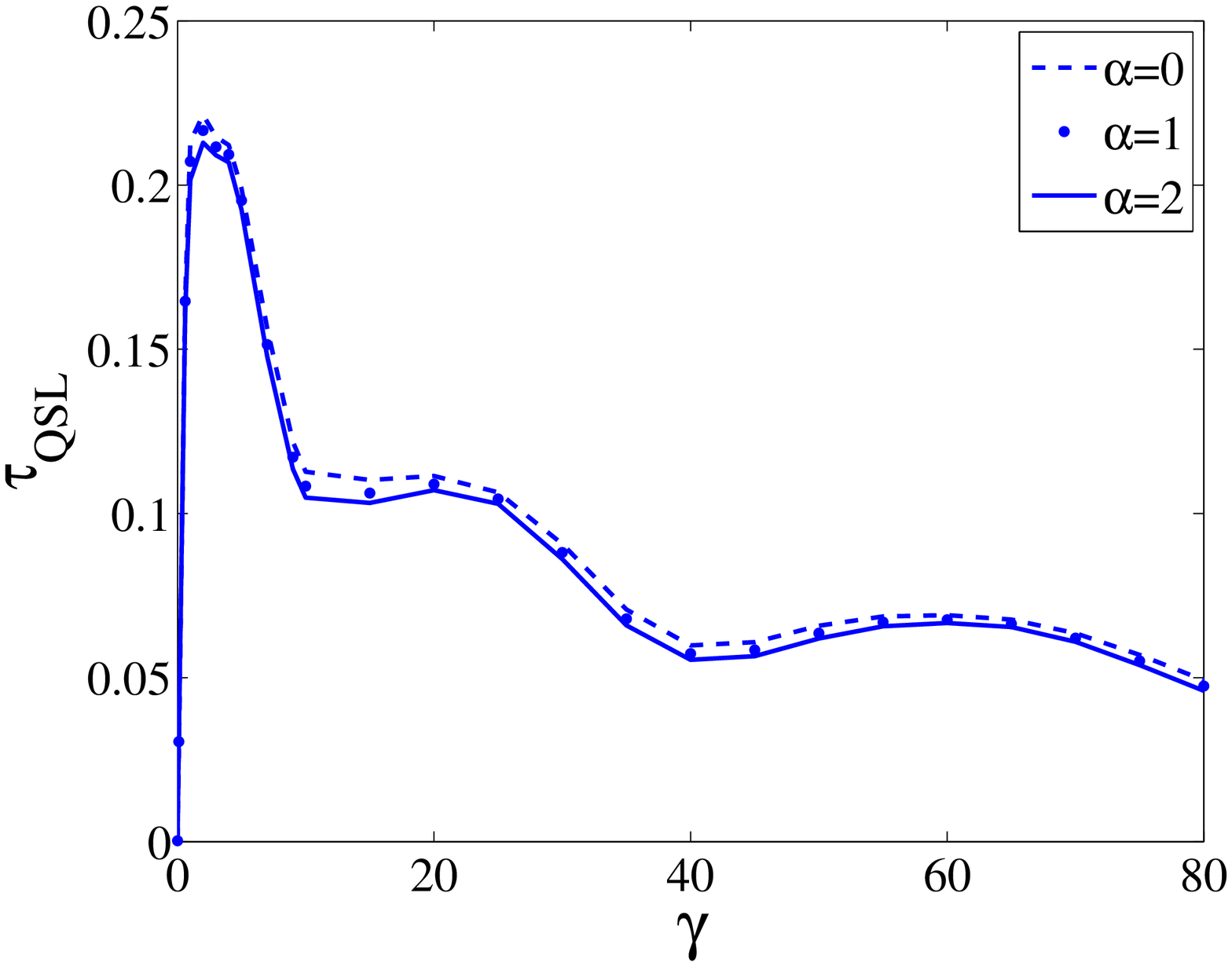}
\caption{} \label{Fig1}
\end{figure}

\newpage
Fig. 11. QSL time versus $\gamma$ with $\theta=1$ and $\lambda=10$. The Horodecki state in (32) is the initial state where $\alpha$ is chosen in the entangled region (free and PPT entangled region). The QSL time for this state relies between the corresponding cases of Werner states presented in Fig. 5 and Fig. 8.
\begin{figure}
\centering
\includegraphics[width=445 pt]{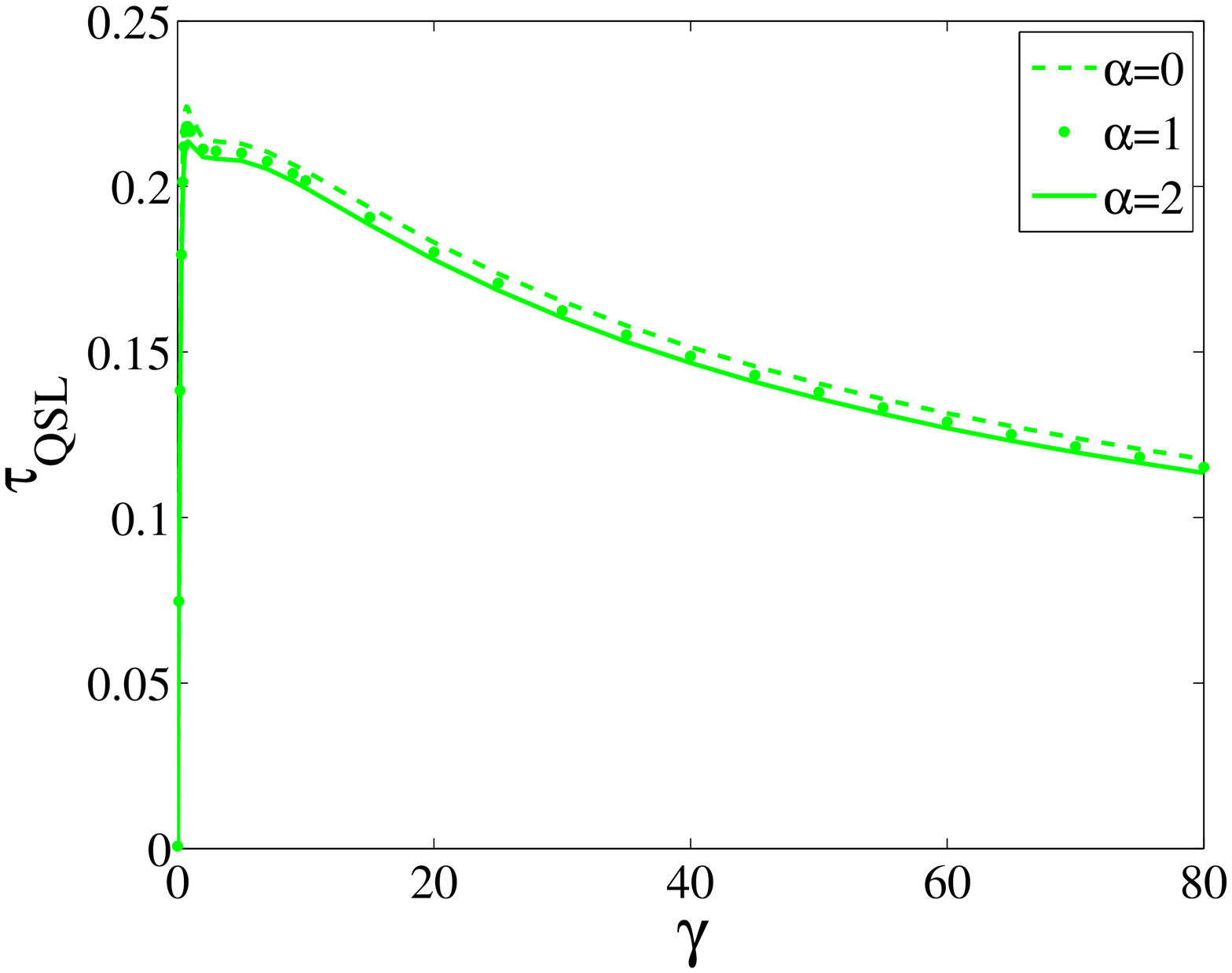}
\caption{} \label{Fig1}
\end{figure}


\newpage
\begin{thebibliography}{99}

\bibitem{Lutz}
S. Deffner and E. Lutz, Phys. Rev. Lett. 105, 170402
(2010).

\bibitem{Giovannetti}
V. Giovannetti, S. Lloyd, and L. Maccone, Nat. Photonics 5, 222 (2011).

\bibitem{Alipour}
S. Alipour, M. Mehboudi, and A. T. Rezakhani, Phys. Rev. Lett. 112, 120405 (2014).

\bibitem{Chin}
A. W. Chin, S. F. Huelga, and M. B. Plenio, Phys. Rev. Lett. 109, 233601 (2012).

\bibitem{Tsang}
M. Tsang, New J. Phys. 15, 073005 (2013).

\bibitem{Caneva}
T. Caneva, M. Murphy, T. Calarco, R. Fazio, S. Montangero, V. Giovannetti, and G. E. Santoro, Phys. Rev.
Lett. 103, 240501 (2009).

\bibitem{Hegerfeldt1}
G. C. Hegerfeldt, Phys. Rev. Lett. 111, 260501 (2013).

\bibitem{Hegerfeldt2}
G. C. Hegerfeldt, Phys. Rev. A 90, 032110 (2014).

\bibitem{Montangero}
S. Lloyd and S. Montangero, Phys. Rev. Lett. 113,
010502 (2014).

\bibitem{Gajdacz}
M. Gajdacz, K. K. Das, J. Arlt, J. F. Sherson, and T.
Opatrn´y, Phys. Rev. A 92, 062106 (2015).

\bibitem{Mukherjee}
V. Mukherjee, A. Carlini, A. Mari, T. Caneva, S. Montangero, T. Calarco, R. Fazio, and V. Giovannetti, Phys.
Rev. A 88, 062326 (2013).

\bibitem{Bekenstein}
J. D. Bekenstein, Phys. Rev. Lett. 46, 623 (1981).

\bibitem{Lloyd}
S. Lloyd, Phys. Rev. Lett. 88, 237901 (2002).

\bibitem{Levitin}
L. B. Levitin, Int. J. Theor. Phys. 21, 299 (1982).

\bibitem{Yung}
M.-H. Yung, Phys. Rev. A 74, 030303 (2006).

\bibitem{Mandelstam}
L. Mandelstam and I. Tamm, J. Phys. (USSR) 9, 249 (1945).

\bibitem{Margolus}
N. Margolus and L. B. Levitin, Physica D 120, 188 (1998).

\bibitem{Taddei}
M. M. Taddei, B. M. Escher, L. Davidovich, and R. L. de Matos Filho, Phys. Rev. Lett. 110, 050402 (2013).

\bibitem{Campo}
A. del Campo, I. L. Egusquiza, M. B. Plenio, and S. F. Huelga, Phys. Rev. Lett. 110, 050403 (2013).

\bibitem{Deffner}
S. Deffner and E. Lutz, Phys. Rev. Lett. 111, 010402 (2013).

\bibitem{Sun}
Z. Sun, J. Liu, J. Ma, and X. Wang, Sci. Rep. 5, 8444 (2015).

\bibitem{che}
Chen Liu, Zhen-Yu Xu, and Shiqun Zhu, Phys. Rev. A 91, 022102 (2015).

\bibitem{Horodecki3}
M. Horodecki, P. Horodecki and R. Horodecki, Phys. Rev. A 60, 1888 (1999).

\bibitem{Horodecki}
P. Horodecki, M. Horodecki and R. Horodecki, Phys. Rev. Lett. 82, 1056 (1999).

\bibitem{Breuer}
H. P. Breuer, E. M. Laine, and J. Piilo, Phys. Rev. Lett. 103, 210401 (2009).

\bibitem{Wen}
Wen-ju Gu and Gao-xiang Li, Phys. Rev. A 85, 014101 (2012).

\bibitem{Wa}
Xiaoguang Wang, Chang-Shui Yu, X.X. Yi, Phys. Lett A 373, 58 (2008).

\bibitem{Horodecki2}
M. Horodecki and P. Horodecki, Phys. Rev. A59, 4206(1999).

\bibitem{Nag}
M.A. Jafarizadeh, N. Behzadi and Y. Akbari, Eur. Phys. J. D 55, 197 (2009).

\end{thebibliography}
\end{document}